\def\year{2020}\relax
\documentclass[letterpaper]{article} 
\usepackage{aaai20}  
\usepackage{times}  
\usepackage{helvet} 
\usepackage{courier}  
\usepackage[hyphens]{url}  
\usepackage{graphicx} 
\usepackage{amsmath}
\usepackage{amssymb}
\usepackage{graphicx}  
\usepackage{subfig}
\usepackage[ruled, linesnumbered]{algorithm2e}
\usepackage{multirow}

\title{High-Order Residual Network for Light Field Super-Resolution}

\author{
Nan Meng\textsuperscript{\rm 1}\thanks{This work was done during an internship at Huawei.}, Xiaofei Wu\textsuperscript{\rm 2}, Jianzhuang Liu\textsuperscript{\rm 2}, Edmund Y. Lam\textsuperscript{\rm 1}\\ 
\textsuperscript{\rm 1}Dept. of Electrical and Electronic Engineering, The University of Hong Kong, Hong Kong\\ 
\textsuperscript{\rm 2}Huawei Noah's Ark Lab, China\\
nanmeng@eee.hku.hk, wuxiaofei2@huawei.com, liu.jianzhuang@huawei.com, elam@eee.hku.hk
}

\begin{document}

\maketitle


\begin{abstract}
Plenoptic cameras usually sacrifice the spatial resolution of their SAIs to acquire geometry information from different viewpoints. Several methods have been proposed to mitigate such spatio-angular trade-off, but seldom make use of the structural properties of the light field (LF) data efficiently. In this paper, we propose a novel high-order residual network to learn the geometric features hierarchically from the LF for reconstruction. An important component in the proposed network is the high-order residual block (HRB), which learns the local geometric features by considering the information from all input views. After fully obtaining the local features learned from each HRB, our model extracts the representative geometric features for spatio-angular upsampling through the global residual learning. Additionally, a refinement network is followed to further enhance the spatial details by minimizing a perceptual loss. Compared with previous work, our model is tailored to the rich structure inherent in the LF, and therefore can reduce the artifacts near non-Lambertian and occlusion regions. Experimental results show that our approach enables high-quality reconstruction even in challenging regions and outperforms state-of-the-art single image or LF reconstruction methods with both quantitative measurements and visual evaluation.
\end{abstract}

\noindent 
Compared to a 2D imaging system, the plenoptic camera not only captures the accumulated intensity of a light ray at each point in space, but also provides the directional radiance information. Together they form the light field (LF), which has shown advantages over 2D imagery in problems such as disparity estimation~\cite{Jeon2015Accurate,Sun2016Data} or 3D reconstruction~\cite{Heber2017Neural} of a scene, generation of images for a novel viewpoint~\cite{Kalantari2016Learning,Meng2019Computational}, and refocusing~\cite{Mitra2012Light}.

Nevertheless, in practice, it can be difficult to achieve a dense sampling of the entire LF due to the limited resolution of the camera sensor. Acquisition of a densely sampled sub-aperture image (SAI) usually sacrifices the view point information, or vice versa~\cite{Wanner2014Variational}. As a result, the LF views exhibit a lower spatial resolution than images obtained by conventional cameras, and many applications such as depth estimation are constrained by low-resolution (LR) images, which increases the significance of the algorithms for light field super-resolution (LFSR). 

Different from single image super-resolution (SISR), a LF is characterized by a structure that needs to be maintained when increasing the data resolution. Such structural information is implicitly encoded in the neighboring views, leading to non-integral shift between two corresponding pixels in the view images~\cite{Wu2017LightTIP}. From this perspective, most depth-based approaches~\cite{Wanner2014Variational,Mitra2012Light} generally depend on such geometric properties as priors to explicitly register the novel SAIs from other views. Their success depend on the accurate depth information, which however is challenging to acquire. Consequently, the disparity errors give rise to artifacts such as tearing and ghosting, especially in the occluded or non-Lambertian areas where depth information is not properly estimated.

Recently, deep learning has shown to be powerful in various computer vision applications~\cite{Meng2018Large,Yang2019Triplet}, including LFSR~\cite{Wu2017Light,Meng2019Spatial,Meng2019High}. The learning-based approaches relieve the dependency on explicit depth information, leading to the improvement of robustness at depth discontinuities. However, the intrinsic limitation of 2D (or 3D) convolution makes existing frameworks difficult to handle the high-dimensional structure in a LF, and therefore most learning-based algorithms simplify the reconstruction to consider only the spatio-angular relations in epipolar plane images (EPIs)~\cite{Wu2017Light}, or the angular correlations among adjacent views~\cite{Yoon2017Light,Zhang2019Residual}. Given that the geometric information is encoded in a complex way within the LF, such simplifications result in performance degradation.

To remedy the problems of existing learning-based approaches for LFSR, we propose to establish a framework tailored to the LF structural information. Such an approach enables the network to learn representations by fully exploiting the LF information from all adjacent views. The main contributions of our model are threefold:
1) We propose a novel high-order structure, named high-order residual block (HRB) to learn the features by fully considering the information from all SAIs of a LF. Such features extracted from HRB preserve high angular coherence.
2) By stacking a set of the HRBs, the proposed network is able to extract diverse spatial features endowed with  scene geometry information. 
In addition, the network also propagates the geometric information encoded in the learned features to achieve high reconstruction quality with the LF structural property.
3) Experimental results demonstrate that our model not only outperforms the state-of-the-art reconstruction methods on quantitative measurements but also generates spatial details and novel views with better fidelity.

\section{Related Work}

\subsection{Spatial super-resolution}
A number of super-resolution algorithms~\cite{Bishop2012Light,Lim2009Improving,Vagharshakyan2018Light} have been developed specifically for LF data. For example, in~\cite{Wanner2014Variational}, a variational framework is applied to super-resolve a novel view based on a continuous depth map calculated from the epipolar plane images (EPIs). Mitra and Veeraraghavan~\cite{Mitra2012Light} proposed a patch-based approach, which is also based on the estimated depth information. These 
methods usually require an accurate  estimate of the disparity information, which can be challenging on LR images. Learning-based approaches mitigate the dependency on geometric disparity, and therefore are more robust in regions where the depth information is difficult to estimate correctly. In~\cite{Yoon2017Light}, LFCNN cascades two CNNs to enhance the target views and generate novel perspectives based on the super-resolved views. However, the stepwise processing does not make use of the entire structural information of the LF, and therefore limits the potential of the model. Recently,~\cite{Wang2018Lfnet} employed a bidirectional recurrent CNN framework to model the spatial correlations horizontally and vertically. Likewise,~\cite{Farrugia2017Super} adopted an example-based spatial SR algorithm on the patch-volumes across the SAIs. Both approaches consider the LF as an image sequence, and therefore lose one angular dimension information.~\cite{Zhang2019Residual}, however, utilized the relations of SAIs in 4 different directions to super-resolve the center target image.
By considering the angular information from multiple directions, their model has superior performance over other previous methods.

\subsection{Angular super-resolution}
Angular super-resolution for LF is also known as view synthesis. Many techniques~\cite{Kalantari2016Learning,Meng2019Computational,Wanner2014Variational} take advantage of the disparity map to warp the existing SAIs to novel views. For instance, \cite{Pearson2013Plenoptic} introduced a layer-based synthesis method to render arbitrary views by using probabilistic interpolation and calculating the depth layer information. \cite{Zhang2015Light} adopted a phase-based method, which integrates the disparity into a phase term of a reference image to warp the input view to any close novel view.

Similar to spatial super-resolution, depth-based techniques are inadequate in the occluded and textureless regions, which prompts researchers to explore algorithms based on CNNs. \cite{Flynn2016Deepstereo} are among the first to apply deep learning to view synthesis from a set of images with wide baselines. Meanwhile,~\cite{Kalantari2016Learning} exploited two sequential CNNs to estimate depth and color information, and subsequently wrapped them to generate the novel view. The dependency on disparity restricts the model performance and easily results in ghosting effects near occluded regions. In~\cite{Wu2017Light}, the author proposed to use a blur-deblur scheme to address the asymmetry problem caused by sparse angular sampling. However, this EPI-based model only utilizes horizontal or vertical angular correlations of a low-resolution LF, which severely restricts the accessible information of the model. Recently, \cite{Yeung2018Fast} applied the alternating convolution to learn the spatio-angular clues for view synthesis and achieves more accurate results.

Compared with the aforementioned approaches, we explore a deeper residual structure for both spatial and angular SR of the LF. The proposed network can harness the high-dimensional LF data efficiently to extract geometric features, which contribute to the high reconstruction accuracy.


\section{Method}
\subsection{Problem formulation}
Following~\cite{Levoy1996Light,Gortler1996Lumigraph}, a light ray is defined by the intersection points of an angular plane $(s, t)$ and a spatial plane $(x, y)$. We consider the LFSR as the recovery of the HR LF $I^H(x,y,s,t) \in R^{\gamma_s X\times \gamma_s Y\times \gamma_a S\times \gamma_a T}$ from the input LR LF $I^L(x,y,s,t) \in R^{X\times Y\times S\times T}$ by two spatial and angular SR factors $\gamma_s$ and $\gamma_a$, respectively. The learning-based SR process can be described as
\begin{equation}
I^S(x,y,s,t) = g(I^L(x,y,s,t); \Theta),
\end{equation}
where $I^S$ stands for the super-resolved LF, $g(\cdot)$ represents the mapping from LR to super-resolved LF, and $\Theta$ denotes the parameters of the model. 

\begin{figure*}[t]
    \centering
    \includegraphics[width=1.\textwidth]{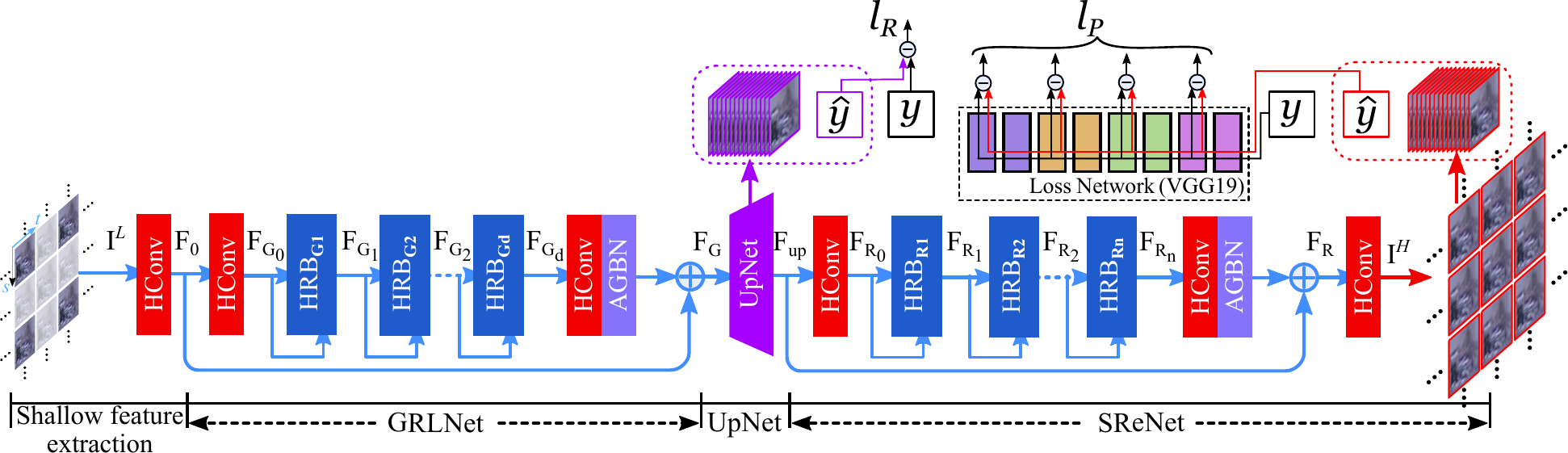}
    \caption{The architecture of our proposed hierarchical high-order network.}
    \label{fig:framework}
\end{figure*}

\subsection{Architecture overview}
The intrinsic limitation of 2D and 3D convolution makes existing schemes unable to fully exploit highly-correlated 4D LF data. As a result, most existing methods consider only partial spatio-angular relations (e.g., EPI)~\cite{Wu2017Light,Wu2018Light}, or angular correlations (e.g., SAI)~\cite{Kalantari2016Learning,Wang2018Lfnet,Zhang2019Residual} which underuse the potential of LF. To resolve the problem, we employ a high-order convolution (HConv) that encapsulates the information from all coordinates by convolving a 4D kernel with the inputs. For any hidden layer $\mathbf{H}^{(k)} \; (k \in \{1,2,\dots,K\})$, the operation of the HConv (together with the following activation layer) is implemented as $\mathbf{H}^{(k)} = \delta(\mathbf{W}^{k} \star \mathbf{H}^{(k-1)})$. $\mathbf{W}^{(k)}$ denotes the weights of the $k^\mathrm{th}$ layer with size $s_1 \times s_2 \times a_1 \times a_2 \times c$, where $c$ is the channel number of the filter bank, $s_1 \times s_2$ is the spatial filter size and $a_1 \times a_2$ is the angular filter size. $\mathbf{H}^{(0)}$ stands for the input $I^L$, and the activation function $\delta(\cdot)$ is the leaky rectified linear unit (LReLU) with slope $\alpha=0.2$. The notation $\star$ is the convolution between an input feature map and a filter.
Furthermore, to utilize spatial information hierarchically from all SAIs, we design the high-order residual block (HRB) to effectively extract the \emph{geometric features} from a LF.
As is shown in Fig.~\ref{fig:framework}, the proposed network mainly consists of four parts: 1) shallow feature extraction, 2) geometric representation learning network (GRLNet), 3) upsampling network (UpNet), and 4) spatial refinement network (SReNet). Specifically, we use a HConv layer to extract shallow features $F_0$ from the LR input:
\begin{equation}
F_0 = H_{\mathrm{HC}}(I^L),
\end{equation}
where $H_{\mathrm{HC}}(\cdot)$ denotes the HConv operation. Subsequently, in the GRLNet, the representations are learned in a hierarchical manner by a set of HRBs. Assuming there are $d$ HRBs, the feature maps $F_{\mathrm{G_d}}$ of the $d^{\mathrm{th}}$ HRB can be expressed as:
\begin{equation}
\begin{aligned}
F_\mathrm{G_d} &= H_{\mathrm{HRB}}^{\mathrm{d}}(F_{\mathrm{G_{d-1}}}) \\
&= H_{\mathrm{HRB}}^{\mathrm{d}}(H_{\mathrm{HRB}}^{\mathrm{d-1}}(F_{\mathrm{G_{d-2}}})) \\
&= H_{\mathrm{HRB}}^{\mathrm{d}} \circ H_{\mathrm{HRB}}^{\mathrm{d-1}} \circ \dots \circ H_{\mathrm{HRB}}^{\mathrm{1}} (F_\mathrm{G_0}),
\end{aligned}
\end{equation}
where $H_{\mathrm{HRB}}^{\mathrm{d}}(\cdot)$ denotes the operation of the $d^{\mathrm{th}}$ HRB, and the symbol $\circ$ stands for function composition. The mapping $H_{\mathrm{HRB}}^{\mathrm{i}}(\cdot), (i=1,2,\dots,d)$ can be a composite function of operations, including $2$ HConvs to fully utilize all the view information within the block (Fig.~\ref{fig:HRB}) to obtain the local geometric feature $F_\mathrm{G_i}$. By cascading multiple HRBs, the geometric features are learned in an hierarchical manner during the training of GRLNet, and therefore more representative features with diverse spatial representations are obtained.
We then apply the global residual learning to combine the hierarchically learned geometric features $F_\mathrm{G_d}$ and the shallow features $F_0$ before conducting upsampling by
\begin{equation}\label{equ:global_residual}
F_\mathrm{G} = H_{\mathrm{AGBN}} \circ H_{\mathrm{HC}}(F_\mathrm{G_d}) + F_0,
\end{equation}
where $H_{\mathrm{AGBN}}$ is an operation of batch normalization defined later.
The following UpNet then upsamples the obtained feature maps $F_\mathrm{G}$ from the LR space to the HR space:

\begin{figure}[!t]
\centering
\subfloat[The angular receptive field of HRB]{
	\includegraphics[width=0.48\columnwidth,height=0.16\textheight]{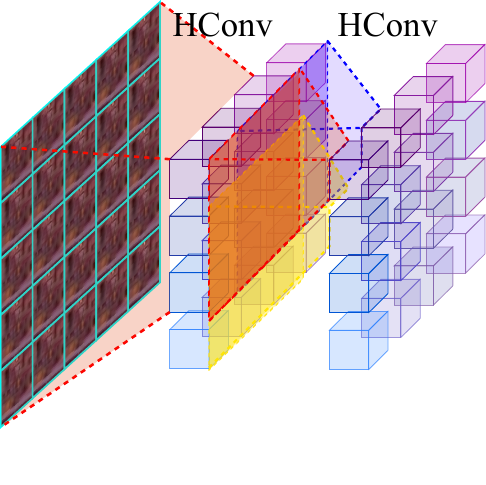}\label{fig:4dconv}
	}\hspace{1mm}
\subfloat[The structure of HRB]{
	\includegraphics[width=0.39\columnwidth]{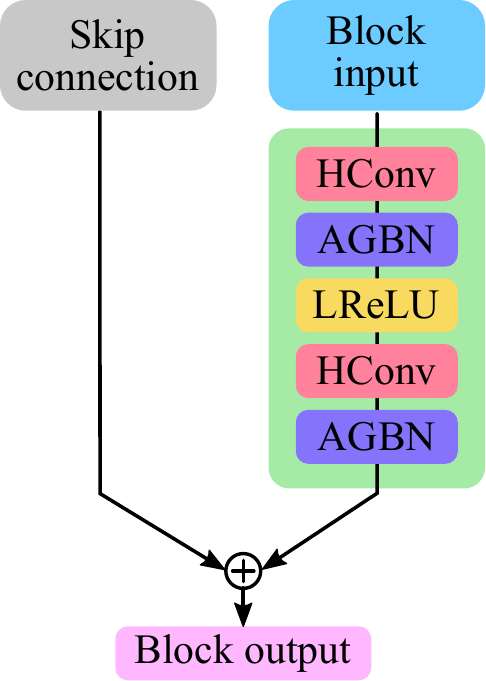}\label{fig:4dresblock}
	}
\caption{The high-order residual block (HRB) architecture, where $\oplus$ denotes the element-wise addition.}
\label{fig:HRB}
\end{figure}

\begin{equation}
F_{\mathrm{up}} = H_{\mathrm{up}}(F_{\mathrm{G}}),
\end{equation}
where $H_\mathrm{up}$ is used to describe the upsampling operation on the LR features. In the experiments, however, directly reconstructing the HR LF based on the fused features is hard, and the results always lack high-frequency spatial details. Therefore, we employ a refinement network (SReNet) supervised by a perceptual loss to recover the spatial details in the HR space:
\begin{equation}
\begin{aligned}
F_{\mathrm{R}} &= H_{\mathrm{AGBN}} \circ H_{\mathrm{HC}}(F_\mathrm{R_n}) + F_{\mathrm{up}},
\end{aligned}
\end{equation}
where $n \leq d$ and $F_{\mathrm{R}}$ denotes the fused refined feature, which is further used for the reconstruction of the final super-resolved LF:
\begin{equation}
I^H = H_\mathrm{HC}(F_{\mathrm{R}}).
\end{equation}
In the following sections, we will illustrate the components of the proposed high-order network in details, and demonstrate the properties of learned geometric features.

\subsection{High-order residual block}

\begin{figure}[!t]
\centering
\subfloat[2D feature slices]{
	\includegraphics[width=0.45\columnwidth]{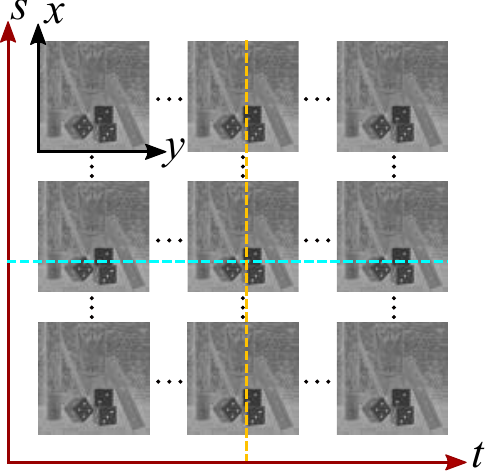}\label{fig:2dslice}
	}
\subfloat[Feature EPIs]{
	\includegraphics[width=0.45\columnwidth]{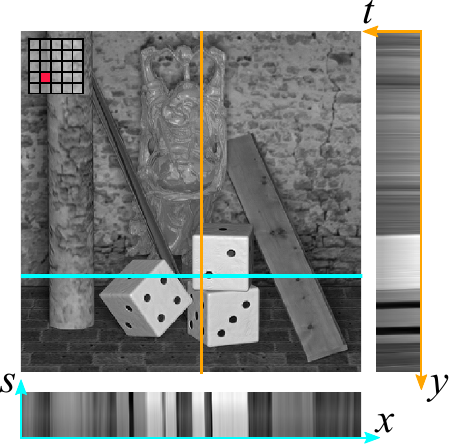}\label{fig:geofea_epi}
	}
	\caption{Visualization of the geometric features. (a) The collection of 2D slices through the learned feature maps. (b) The EPI located at the corresponding lines.}\label{fig:geometric_features}
\end{figure}
As a basic building block in our network, the HRB's structure is presented in Fig.~\ref{fig:HRB}.
According to the Fig.~\ref{fig:4dconv}, each HRB contains two HConv layers with the $3\times3$ angular receptive field which makes the block possible to fully utilize the information from all SAIs of input features. In addition, to ease the training of the proposed high-order network, we apply the normalization operation to the outputs of the HConv layer~\cite{Ioffe2015Batch}. Given that the inputs preserve high coherence among the views, the normalization should not be counted in an aperture-wise manner to avoid that the whitening decorrelates the coherence. We consequently implement the normalization over a group of SAIs in every channel of the feature maps and therefore propose an aperture group batch normalization (AGBN).

Let the outputs of a particular hidden HConv layer be $\mathbf{H}=\{\mathcal{H}_m^n(s,t)\}$, where $s$ and $t$ stand for the two indices of angular dimensions as is defined in the problem formulation. The superscript $n \in [1,N]$ denotes the number of channels, and for each feature SAI contains $M$ values ($m = 1, 2, \dots, M$). Therefore, the AGBN transform is implemented as in Algorithm~\ref{algo:AGBN}.
\begin{algorithm}
\KwIn{Features from HConv layer: $\mathbf{H}=\{\mathcal{H}^n(s,t)\}$; \ Parameters $\gamma$ and $\beta$}
\KwOut{The output features: $\mathbf{\hat{H}}=\{\mathcal{\hat{H}}^n(s,t)\}$}
\For{$n = 1,2,\dots, N$}{
$\; \mu_n \leftarrow \frac{1}{M}\sum_{m=1}^M (\frac{1}{S \cdot T} \sum_{s=1}^S \sum_{t=1}^T \mathcal{H}^n_m(s,t))$\
$\hspace*{1.5em} = \frac{1}{MST} \sum_{m=1}^M \sum_{s=1}^S \sum_{t=1}^T \mathcal{H}^n_m(s,t))$;\
$\sigma_n \leftarrow \frac{1}{MST} \sum_{m=1}^M \sum_{s=1}^S \sum_{t=1}^T (\mathcal{H}^n_m(s,t)-\mu_n)^2$;\
$\mathcal{\hat{H}}^n(s,t) \leftarrow \gamma \cdot \frac{\mathcal{H}^n(s,t) - \mu_n}{\sqrt{\sigma_n^2+\epsilon}} + \beta$
}
\caption{Aperture group batch normalization}\label{algo:AGBN}
\end{algorithm}

By stacking the layers as is shown in Fig.~\ref{fig:4dresblock}, the HRB is able to extract features that preserve geometrical properties by considering all SAIs. The learned geometrical features not only contain spatial structures (such as textures or edges) but also record the relations between adjacent feature views. Fig.~\ref{fig:geometric_features} exhibits an example of the geometric features learned by the HRB. To illustrate such high-dimensional features, we display a grid of 2D slices through the 4D features in Fig.~\ref{fig:2dslice}, and the EPIs located at the corresponding lines in a certain feature slice in Fig.~\ref{fig:geofea_epi}. The feature EPIs very much resemble the LF EPIs, reflecting that the HRB has the capacity to extract features preserving high coherence.

\subsection{Geometric representation learning network}
\begin{figure}[!t]
    \centering
    \includegraphics{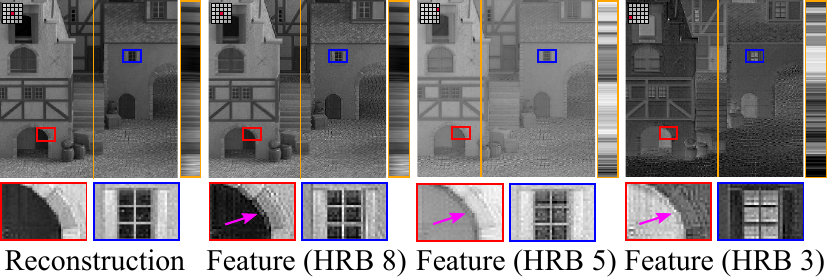}
    \caption{Illustration of the geometric features extracted from different HRBs in the GRLNet.}
    \label{fig:HRB_geofeatures}
\end{figure}
The GRLNet is composed of a set of cascaded HRBs. Such structure enables the network to learn multiple spatial representations endowed with geometry information. Compared with the features extracted from traditional CNN-based model, the learned geometric features are different in two aspects: 1) the high coherence among SAIs in angular dimension; 2) the \emph{smooth} effects near object borders in spatial dimension. The former has been discussed in Fig.~\ref{fig:geofea_epi}, where we show the \emph{EPI property} of features. The latter can be illustrated according to the features from different HRBs. In Fig.~\ref{fig:HRB_geofeatures}, we visualize the spatial appearance of the geometric features extracted from the $3^{\mathrm{rd}}$, $5^{\mathrm{th}}$, and $8^{\mathrm{th}}$ HRBs. The red boxes zoom in at the features of object border, while the blue boxes zoom in at the texture features. Compared with the reconstruction result, the edges of the object border in the feature space (red boxes) are not as sharp. Such effects are caused by the rapid changes in the parallax of the object border, and therefore indicate the scene geometric information. In addition to the diverse spatial representations, the feature angular coherence is also maintained (e.g., the EPI in the yellow boxes). Consequently, the GRLNet is able to learn more representative spatial features hierarchically through a set of HRBs and simultaneously propagates the geometric information.

\subsection{Upsampling network}
\begin{figure}[!t]
    \centering
    \includegraphics[width=0.98\columnwidth]{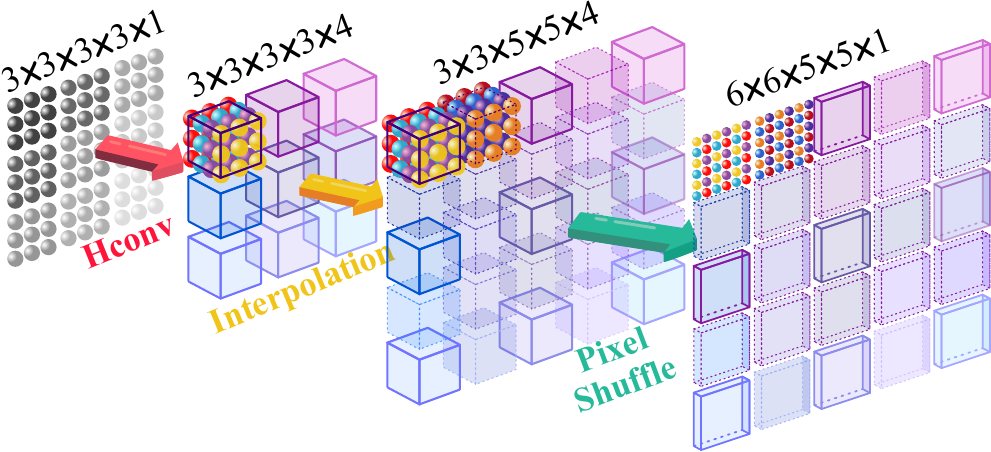}
    \caption{Illustration of UpNet for spatio-angular resolution enhancement. The red arrow stands for HConv opeartion, the yellow one denotes angular linear interpolation, and the green one denotes channel-to-space pixel shuffle operation.}
    \label{fig:upsampling}
\end{figure}
The upsampling network is applied to increase the spatio-angular resolution using the extracted hierarchical geometric features in the LR space. We design the network to fit to the properties of the LF geometric features. As illustrated in Fig.~\ref{fig:upsampling}, we assume a single LR feature map with a single channel as input. The feature map has dimension $X \times Y \times S \times T$, where $X=Y=3$ and $S=T=3$. The spatial and angular upsampling factors are $\gamma_s=\gamma_a=2$ (strictly speaking, the angular dimension is increased from $3\times3$ to $5\times5$). The first step expands the feature channel by a factor of $\gamma_s^2$ using the HConv operation. Then, given the EPI property of the geometric features, we apply a linear interpolation to the angular dimension of the feature maps to upscale the resolution by a factor of $\gamma_a$. Finally, the channel-to-space shuffle operation is applied to increase the spatial resolution by a fator of $\gamma_s$.

The UpNet upsamples the learned geometric features to the HR space. Such features are used to reconstruct the primary super-resolved LF directly to get the per-pixel reconstruction loss for training, and are also passed to the SReNet to further recover the high-frequency details.

\subsection{Spatial refinement network}
The SReNet aims at restoring the realistic spatial details of the previous super-resolved output. Given that the GRLNet is trained using a pixel-wise loss, it tends to generate smooth results with poor fidelity~\cite{Ledig2016Photo,Gupta2011Modified}. The SReNet in contrast learns the geometric features directly in the HR space and is supervised by a novel perceptual loss defined on SAIs to make the reconstruction sharper. In the experiments, we will discuss the effects of the SAI-wise perceptual loss for recovering spatial details. 

\subsection{Loss function}
We propose a two-stage loss function (see Fig.~\ref{fig:framework}) to encourage the proposed network to learn the geometric features and reconstruct high-quality spatial details. In general, the loss function is a linear combination of two terms:
\begin{equation}\label{equ:loss}
\ell = \alpha \cdot \ell_R + \beta \cdot \ell_P.
\end{equation}
The \emph{reconstruction loss} $\ell_R$ models the pixel-wise difference between the super-resolved LF $I^S$ and the ground truth $I^H$:
\begin{equation}\label{equ:reconstruction_loss}
\ell_R = \sum_{x=1}^X \sum_{y=1}^Y \sum_{s=1}^S \sum_{t=1}^T (I^H(x,y,s,t) - I^S(x,y,s,t))^2.
\end{equation}
The \emph{perceptual loss} $\ell_P$ measures the quality of the spatial reconstruction. Inspired by~\cite{Johnson2016Perceptual}, we define the loss function acquired from a VGG network to describe the aperture-wise differenes between high-level features $\phi$,
\begin{equation}
\ell_P = \frac{1}{ST} \sum_{s=1}^S \sum_{t=1}^T \left(\phi(I_{s,t}^H) - \phi \left(g(I_{s,t}^L; \Theta) \right)\right)^2,
\end{equation}
where $I_{s,t}^L = I^L(\cdot,\cdot,s,t)$ and $I_{s,t}^H = I^H(\cdot,\cdot,s,t)$ denote the LR input and the ground truth with angular coordinate $(s,t)$, respectively.

\begin{table*}[!t]
\centering
\caption{Quantitative evaluation of state-of-the-art methods for spatial and angular LFSR. We report the average PSNR and SSIM over all sub-aperture images for Spatial $2\times$, $3\times$, $4\times$ and Angular $3\times$ ($3\times3 \rightarrow 9\times9$). The \textbf{bold} values indicate the best performance.}\label{table:spatialSR}
\includegraphics[width=0.95\textwidth]{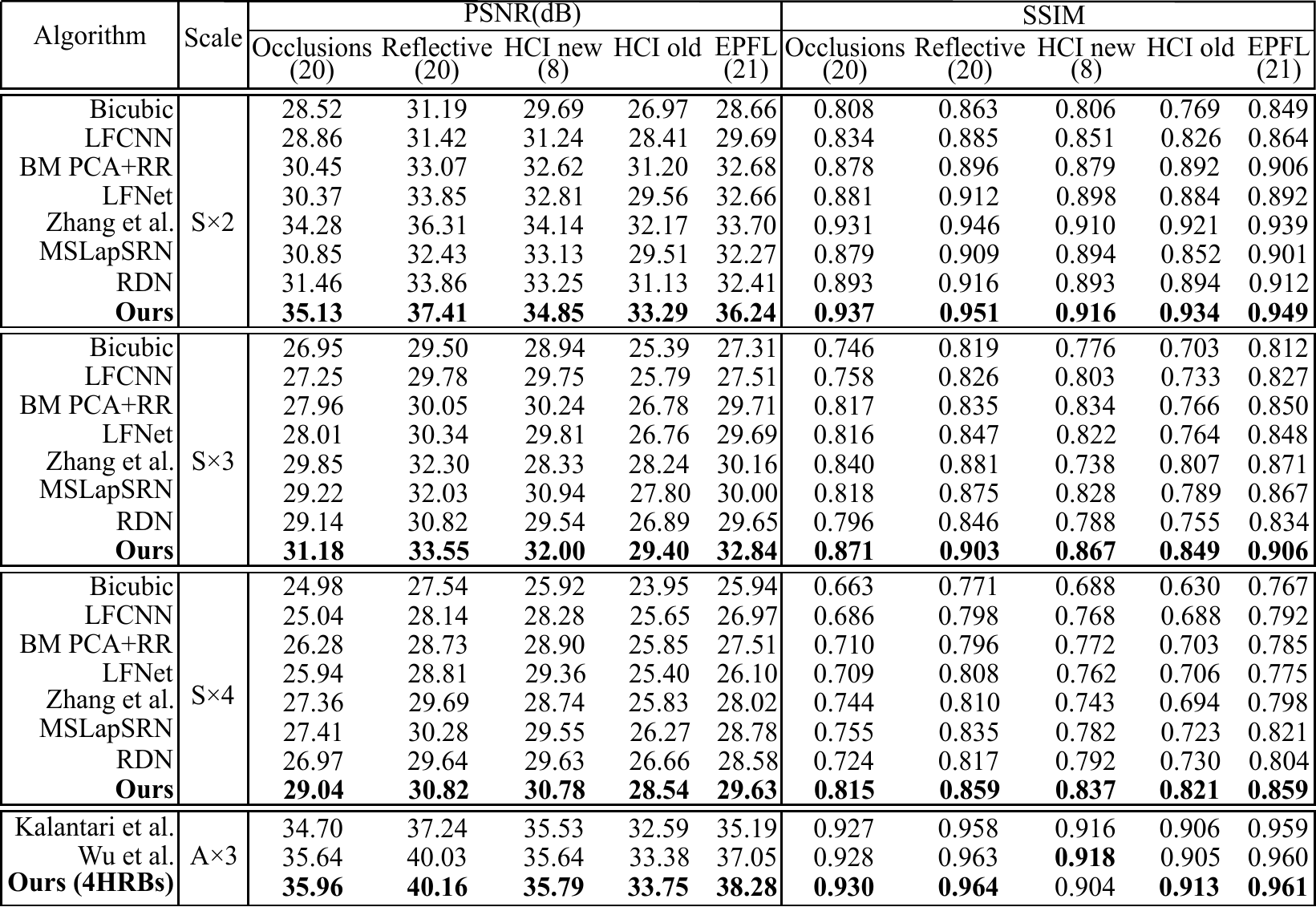}
\end{table*}

\section{Experiments}
\subsection{Data and experiment settings}
In the experiments, we randomly select 100 scenes from the Lytro Archive (Stanford) (excluding ``Occlusions'' and ``Reflective'') and the entire Fraunhofer densely-sampled high-resolution~\cite{Ziegler2017Acquisition} datasets for training. The former consists of 353 real-world scenes captured using a Lytro Illum camera with a small baseline, and in addition, we exclude the corner samples and only select the center $9\times9$ views in the experiments. The latter contains 9 real-world scenes that are densely sampled by a high-resolution camera with a larger baseline. 
The experimental results show that our trained network can be generalized to various synthetic and real-world scenes, as well as some microscopy light fields. This indicates that the learned geometric features are generic for various situations. 

During training, the system each time receives a 4D patch of LF, which is spatially cropped to $96 \times 96$ as input. For spatial SR, the downsampling is based on the classical model~\cite{Farrugia2018Light}
\begin{equation}\label{equ:downsampling}
I^L = \downarrow_{\gamma_s} G * I^H + \eta,
\end{equation}
where $\eta$ is Gaussian noise with zero mean and unit standard deviation, $\downarrow_{\gamma_s}$ denotes the nearest neighbor downsampling operator applied to each view, $\gamma_s$ is the magnification factor, and $G$ stands for a Gaussian blurring kernel with a window size of $7 \times 7$ and standard deviation of $1.2$ pixels. 
The network is trained using the Stochastic Gradient Descent solver with the initial learning rate of $10^{-5}$, which is decreased by a factor of $0.1$ for every 10 epochs. The entire implementation is available at \textit{https://github.com/monaen/LightFieldReconstruction}.

\subsection{Loss evaluation}
To exam the effectiveness of different loss components, we adjust the proposed network to obtain multiple variants which are further trained using different losses. The parameters of different variants are kept constant to control the model representational capacity. In Table~\ref{table:ablation_losses}, we evaluate the performance of the variants with 8 HRBs in total. By combining the reconstruction and perceptual losses, the model can achieve comprehensively better quantitative results and reconstruct the LF with good visual fidelity (refer to the supplementary materials).
\begin{table}[!h]
\centering
\caption{Ablation study of different components in the proposed model. ``G8'' denotes 8 HRBs in GRLNet, ``S8'' denotes 8HRBs in SReNet, and ``G5S3'' stands for 5 HRBs in GRLNet and 3 HRBs in SReNet.}\label{table:ablation_losses}
\includegraphics[width=1.\columnwidth]{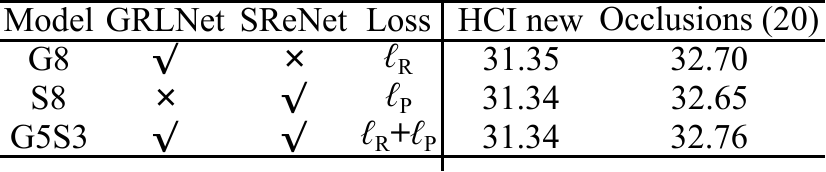}
\end{table}

\subsection{Spatial super-resolution evaluation}
For evaluation in terms of spatial resolution, we compare against several top-performing algorithms designed for LFSR, including LFCNN~\cite{Yoon2017Light}, BM PCA+RR~\cite{Farrugia2017Super}, LFNet~\cite{Wang2018Lfnet}, Zhang et al.~\cite{Zhang2019Residual} and some state-of-the-art methods for SISR, like MSLapSRN~\cite{Lai2018Fast} and RDN~\cite{Zhang2018Residual}. For fair comparison, all the methods are retrained using the same datasets and downsampling method described in Eq.~\ref{equ:downsampling}. Table~\ref{table:spatialSR} shows the quantitative comparisons for $\times2$, $\times3$, and $\times4$ SR on five public LF dataset. The real-world datasets consist of 20 scenes from ``Occlusions'' and 20 scenes from ``Reflective'' in Stanford Lytro Archive (Stanford), and 21 scenes from EPFL~\cite{Rerabek2016New}, while the synthetic datasets are selected from the HCI dataset~\cite{Honauer2016Benchmark,Wanner2013Datasets}. We carefully fine-tune and retrain each algorithm to fit the classic downsampling method described in Eq.~\ref{equ:downsampling} using their publicly available code to reach their best performance.
The results are measured in terms of the peak signal-to-noise ratio (PSNR) and structural similarity (SSIM) over the center $9\times9$ views of each evaluation scene, and we report the average value on each test dataset. 

\begin{figure}[!t]
    \centering
    \includegraphics[height=0.25\textheight]{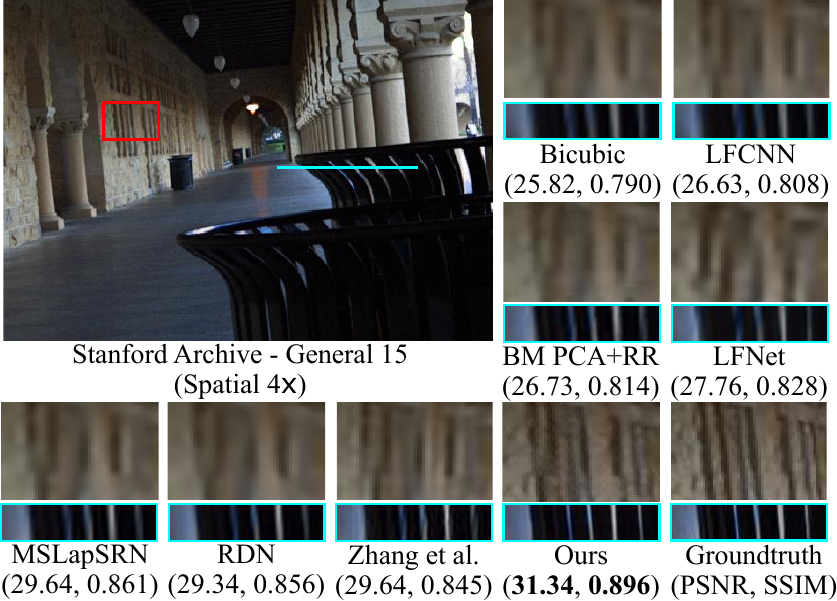}
    \caption{Visual comparison for $4\times$ spatial SR on the real-world scene General 15 from Stanford Archive.}
    \label{fig:real-spatial4x}
\end{figure}

Fig.~\ref{fig:real-spatial4x} compares the visual reconstruction results for $4\times$ spatial SR. For LFSR algorithms, LFCNN receives pairs of SAIs as input without modeling their correlations. Therefore, it underuses the angular information and tends to generate over-smoothed results with blurry details. Likewise, the two SISR methods (MSLapSRN and RDN) do not model the correlations either, which leads to the blurring and twisting in their EPIs.
BM PCA+RR and LFNet simplify the problem by considering only one dimension of angular correlations. Such strategy restricts the performance of their methods in terms of both visual results (Fig.~\ref{fig:real-spatial4x}) and quantitative measurements (Table~\ref{table:spatialSR}). In contrast, a recent approach proposed by Zhang et al.~\cite{Zhang2019Residual} exploits the angular information from 4 directions for LF spatial SR. By integrating more directional information, the algorithm shows better quantitative results on spatial $2\times$ and $3\times$ tasks. Nevertheless, their approach fuses the angular information by roughly concatenating SAIs from $4$ directions in the channel dimension. Given that the differences between adjacent views decrease rapidly in the LR LF, its performance on $4\times$ spatial SR is badly affected. Compared with these methods, our model exploits the entire angular information from all directions. In addition, such angular information is further fused in the learned geometric features during the network training. In this way, all of the structural information is utilized for the final reconstruction allowing our model to achieve superior performance in terms of both visual fidelity (e.g., the texture on the wall in Fig.~\ref{fig:real-spatial4x}) and quantitative measurements. More visual results on both real-world and synthetic scenes are presented in our supplementary materials to illustrate the model generalization ability.

\subsection{Angular super-resolution evaluation}
\begin{figure}[!t]
    \centering
    \includegraphics[width=1.\columnwidth,height=0.11\textheight]{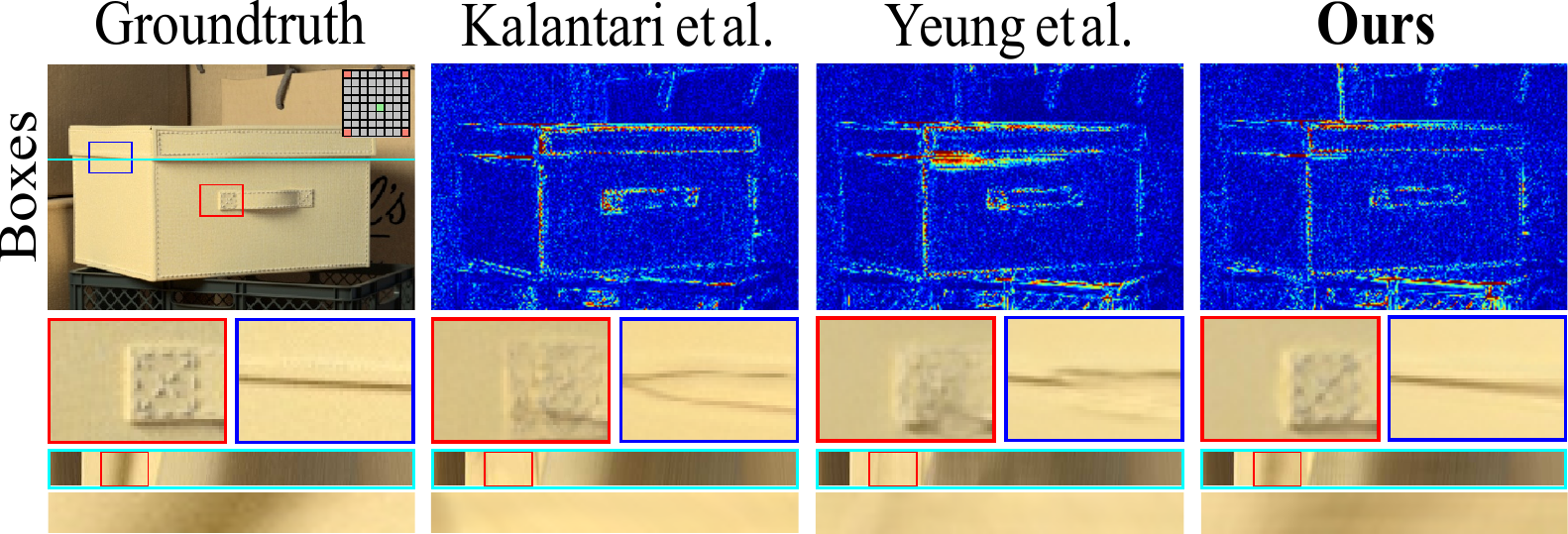}
    \includegraphics[width=1.\columnwidth,height=0.11\textheight]{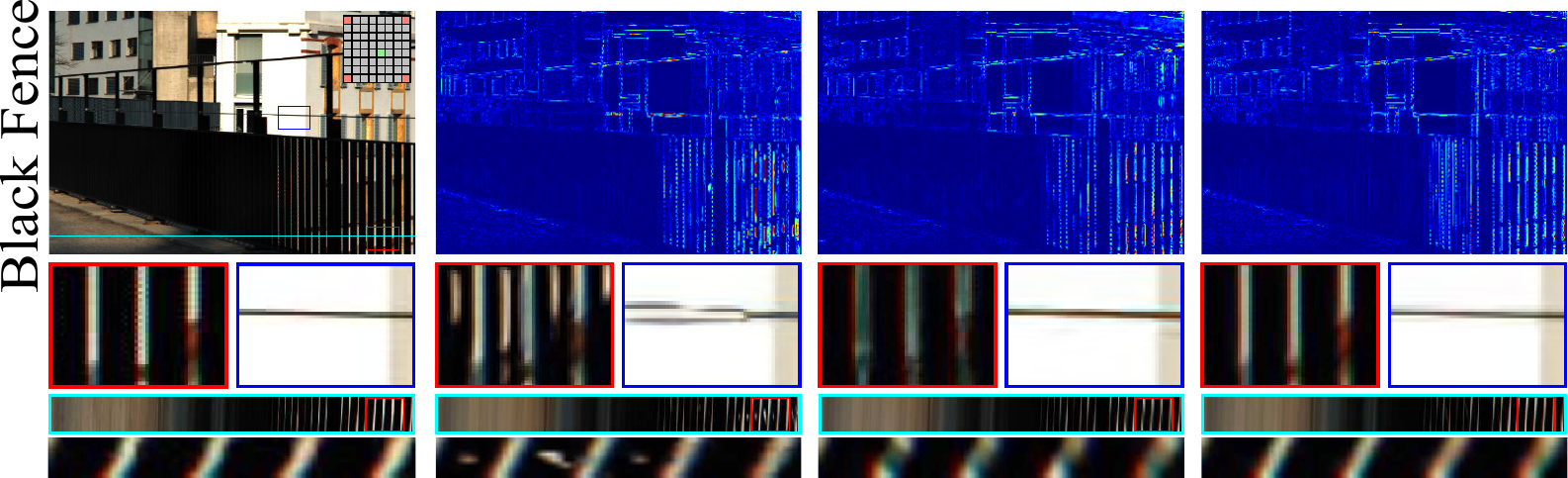}
    \includegraphics[width=1.\columnwidth,height=0.11\textheight]{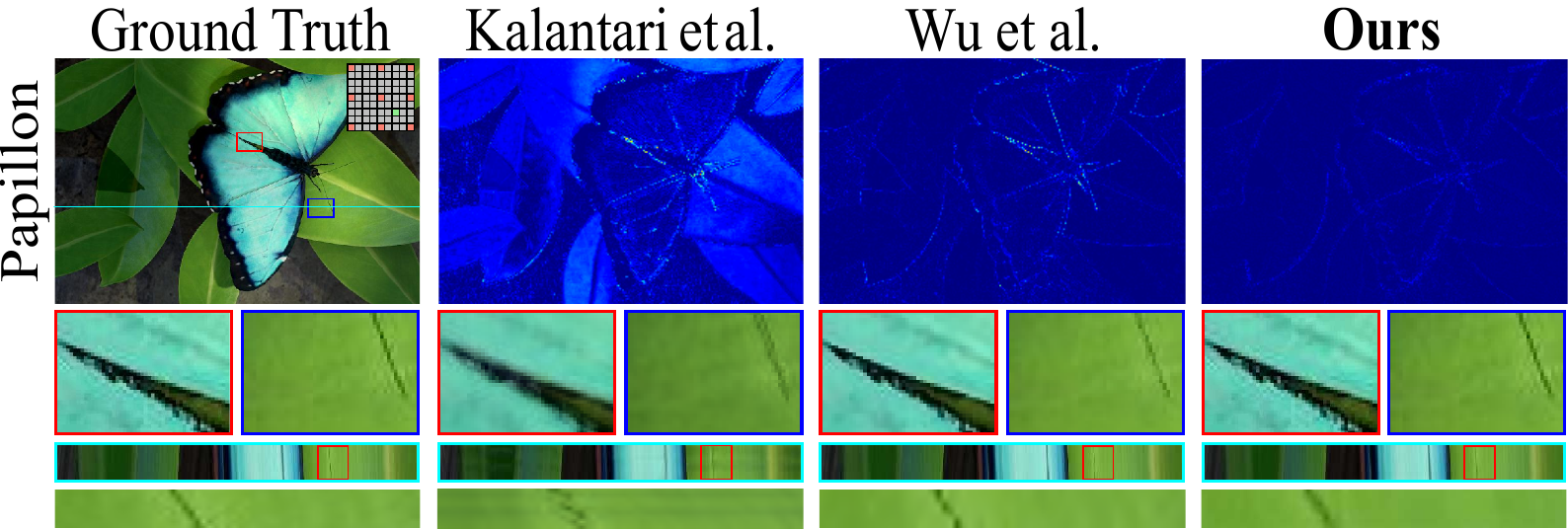}
    \includegraphics[width=1.\columnwidth,height=0.11\textheight]{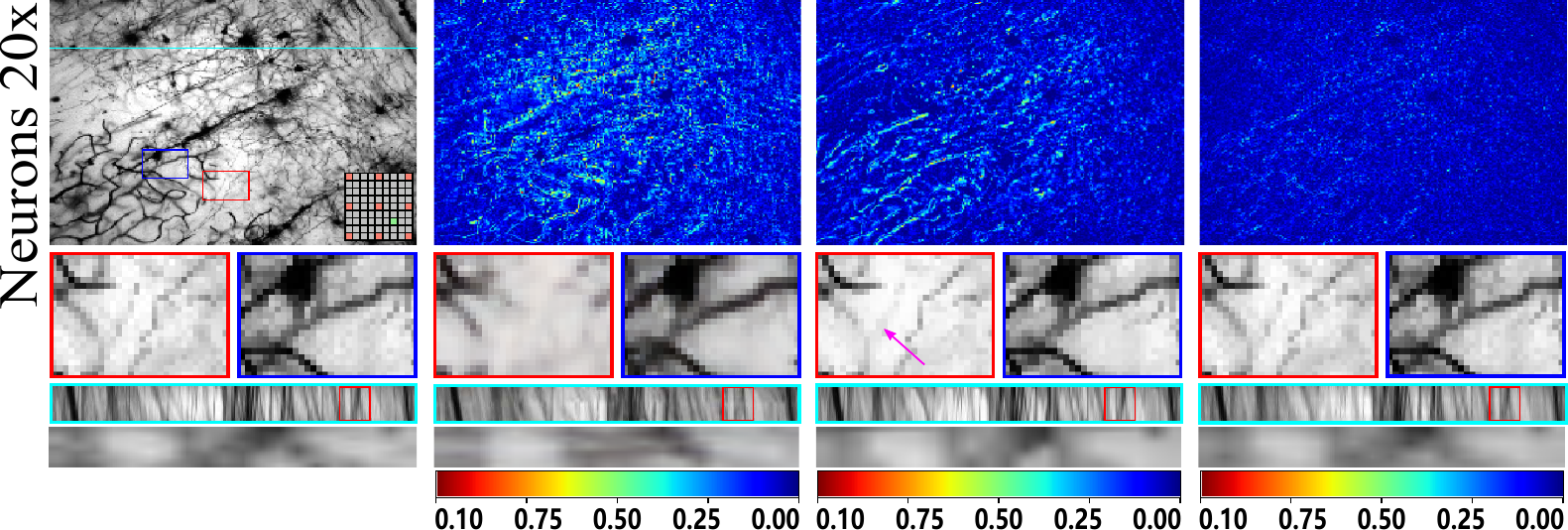}
    \caption{Visual comparison of our model with Kalantari et al. and Yeung et al. for the $2\times2 \rightarrow 8\times8$ task, and with Wu et al. for the $3\times3 \rightarrow 9\times9$ task.}
    \label{fig:angularSR}
\end{figure}
For angular super-resolution, all the models are trained only using the 100 Lytro Archive scenes. We carry out comparisons with three state-of-the-art CNN based methods, namely, Kalantari et al.~\cite{Kalantari2016Learning}, Wu et al.~\cite{Wu2017Light} and Yeung et al.~\cite{Yeung2018Fast}. For all the angular SR task, we evaluate a simplified version of the proposed model with $3$ HRBs in GRLNet and $1$ HRB in SReNet ($8$ HConvs in total). Even with only $4$ HRBs, our model is able to defeat the other three methods.
For the $3\times3 \rightarrow 9\times9$ synthesis task, the quantitative comparisons on average PSNR and SSIM are presented in the last part of Table~\ref{table:spatialSR}. Our model defeats Kalantari et al. and Wu et al. on most real-world and synthetic LF scenes. 
Fig.~\ref{fig:angularSR} compares the visual results. The depth-dependent method Kalantari et al. tends to produce ghosting artifacts near boundaries of objects. Wu et al. only uses the EPI information and therefore leads to a loss of spatial details (e.g., Neurons $20\times$ nerve fibre is absent in their reconstructed LF). To demonstrate the effectiveness of the hierarchical HRBs structures, we further compare the performance against Yeung et al.'s model with 16 4D alternating convolutions (16L). According to the Table~\ref{table:2x2_8x8_angular} and Fig.\ref{fig:angularSR}, our model achieves higher quantitative values and synthesizes more realistic novel views (also refer to our supplementary video).

\begin{table}[!t]
\label{table:2x2_8x8_angular}
\centering
\includegraphics[width=1.\columnwidth]{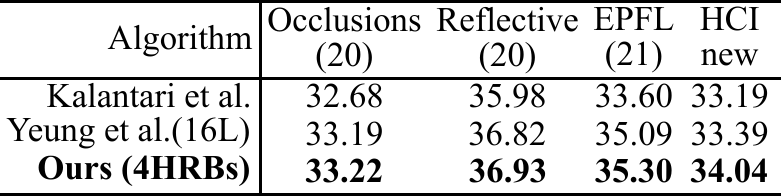}
\caption{Quantitative evaluation of state-of-the-art view synthesis algorithms. We report the average PSNR for the task $2\times2 \rightarrow 8\times8$.}\label{table:2x2_8x8_angular}
\end{table}
\section{Conclusion}
In this paper, we design a hierarchical high-order framework for LF spatial and angular SR. To fully exploit the structural information of the LF, HRB has been proposed. By cascading a set of HRBs, our model is able to extract representative features encoded with geometric information. Such features contribute a lot to the final reconstruction results. In addition, the combination of the pixel-wise loss and the perceptual loss further allows our model to generate more realistic spatial images. The experiments show that our proposed model outperforms the state-of-the-art SR methods in terms of both quantitative measurements and visual fidelity.

\section{Acknowledgments}
This work is supported in part by the Research Grants Council of Hong Kong (GRF 17203217, 17201818, 17200019) and the University of Hong Kong (104005009, 104005438).

\bibliography{aaai2020}
\bibliographystyle{aaai}


\end{document}